\documentclass[utf8]{frontiersSCNS} 

\usepackage{url,hyperref,lineno,microtype,subcaption}
\usepackage[onehalfspacing]{setspace}

\def\keyFont{\fontsize{8}{11}\helveticabold }
\def\firstAuthorLast{Christianson \& Garrod} 
\def\Authors{Drew A. Christianson\,$^{*1}$ and Robin T. Garrod\,$^{1,2}$}

\def\Address{$^{1}$Department of Chemistry, University of Virginia, Charlottesville, VA, USA \\
$^{2}$Department of Astronomy, University of Virginia, Charlottesville, VA, USA  }

\def\corrAuthor{Drew A. Christianson, Dept. of Chemistry, University of Virginia, McCormick Rd, Charlottesville, VA 22904, USA}

\def\corrEmail{dac5tt@virginia.edu}

\begin{document}
\onecolumn
\firstpage{1}

\title[Chemistry on porous interstellar dust]{Chemical kinetics simulations of ice chemistry on porous versus non-porous interstellar dust grains} 

\author[\firstAuthorLast ]{\Authors} 
\address{} 
\correspondance{} 

\extraAuth{}

\maketitle

\begin{abstract}

\section{}
The degree of porosity in interstellar dust-grain material is poorly defined, although recent work has suggested that the grains could be highly porous. Aside from influencing the optical properties of the dust, porosity has the potential to affect the chemistry occurring on dust-grain surfaces, via increased surface area, enhanced local binding energies, and the possibility of trapping of molecules within the pores as ice mantles build up on the grains. 
Through computational kinetics simulations, we investigate how interstellar grain-surface chemistry and ice composition are affected by the porosity of the underlying dust-grain material. Using a simple routine, idealized three-dimensional dust-grains are constructed, atom by atom, with varying degrees of porosity. Diffusive chemistry is then simulated on these surfaces using the off-lattice microscopic Monte Carlo chemical kinetics model, {\em MIMICK}, assuming physical conditions appropriate to dark interstellar clouds. On the porous grain surface, the build-up of ice mantles, mostly composed of water, leads to the covering over of the pores, leaving empty pockets. Once the pores are completely covered, the chemical and structural behavior is similar to non-porous grains of the same size. The most prominent chemical effect of the presence of grain porosity is the trapping of molecular hydrogen, formed on the grain surfaces, within the ices and voids inside the grain pores. Trapping of H$_2$ in this way may indicate that other volatiles, such as inert gases not included in these models, could be trapped within dust-grain porous structures when ices begin to form.

\tiny
 \keyFont{ \section{Keywords:} astrochemistry, chemical kinetic simulations, dark interstellar clouds, dust porosity, interstellar dust, interstellar ice, molecular hydrogen, surface chemistry}
\end{abstract}

\section{Introduction}

Dust grains are known to play a major role in the chemical evolution of interstellar clouds and star-forming regions. Of critical importance is the production of molecular hydrogen on grain surfaces, through the reactions of hydrogen atoms adsorbed from the gas phase \citep[e.g.][]{gouldsalpeter,Hollenbach71}; much of the H$_2$ produced in this way desorbs again, to become the dominant component of the gas in dense, dark clouds. In clouds of sufficient visual extinction, the dust grains may build up ice mantles composed mostly of water, but with substantial amounts of CO, CO$_2$, NH$_3$ and several other simple species \citep{boogert}. Much of the chemistry that produces these grain-surface molecules involves the surface diffusion of atomic hydrogen, which is understood to be mobile even under the cold ($\sim$ 10~K) conditions of dark clouds \citep{Sen17}. By repetitive thermal hopping between surface potential minima (binding sites), H atoms may meet other reactive species, such as atomic O or N, and react. The adsorption of atoms, as well as molecules such as CO, from the gas phase and their subsequent hydrogenation leads to the build up of ice mantles many monolayers in thickness \citep{cuppen}.

Interstellar dust grains are typically considered to be composed of either carbonaceous or silicaceous material, both of which populations may be identified in the interstellar medium. By fitting observational extinction values to models that consider the optical properties of dust grains, \citet{mathis77} provided a grain-size distribution for the Milky Way of $dn \propto a^{-3.5} da$, valid for grains with radii in the range $50$~\AA $< a < 0.25$~$\mu$m. \citet{weingartner} produced more detailed expressions using more recent observational data, providing distinct distributions for carbon versus silicate grains, as well as bringing the minimum size down to molecular scales (i.e. a few \AA). 

In chemical kinetic models of interstellar grain-surface chemistry, in which the grain-surface populations of various atomic and molecular species are traced over astronomical timescales, a canonical, spherical grain radius of 0.1~$\mu$m is typically chosen to represent all grains in the model. The grain surface is assumed to comprise around 1 million binding sites, based on a site density of $\sim 10^{15}$~cm$^{-2}$ \citep{hasegawa}. Some chemical models of interstellar clouds have considered a distribution of grain sizes, each having their own independent chemistry \citep{acharyya,paulygarrod16,paulygarrod18}. However, without the existence of a strong temperature gradient across the size distribution, the influence of the grain size on the surface chemistry in those models appears to be fairly limited.

Most such chemical kinetic models use the so-called ``rate-equation'' approach \citep[see][]{garrod08}, in which the grain-surface population of each chemical species is traced over time; each chemical species is assigned a fixed binding energy and barrier against diffusion. While efficient, this technique is unable to treat the microscopic structure of either the ice mantle or the underlying dust grain itself, as the positions of individual atoms or molecules on the grain are not considered. The dust grains in these models are thus implicitly smooth and spherical, as are the molecular ices that form on their surfaces. Most models therefore say nothing about the possible influence of grain morphology or porosity on interstellar molecular abundances, either on the grain surfaces or in the gas phase.

Experimental investigations of the direct deposition of interstellar ice analogs suggest the possibility for grain-surface ices to achieve substantial degrees of porosity \citep[e.g][]{raut}, but no observational evidence has yet been found to indicate that true interstellar ices, which are formed primarily through surface chemistry rather than direct deposition, should be substantially porous; the dangling bond IR feature associated with porous ASW has not so far been detected in the ISM \citep{keane}. Laboratory work indicates that the formation of ices through chemical reactions, rather than direct deposition from the gas phase, appears to be crucial to the degree of porosity. In experiments by \citet{Oba09}, hydrogen atoms and O$_2$ molecules were co-deposited onto Al and D$_2$O surfaces, leading to the build up of water ice films through chemical reactions. Those authors found no dangling-bond features that would indicate porosity, and they suggested that the reaction energy could be responsible for the compactness of the ice. Similar work by \citet{Accolla13} involving D and O$_2$ deposition likewise found no spectral identifiers of porosity.

Astrochemical models of interstellar ice formation by \citet{garrod13a} used a microscopic Monte Carlo technique to trace the formation of water and several other chemical species through atomic addition reactions. Those authors found that the ability of oxygen atoms to diffuse on the grain/ice surface over long periods of time allowed them to find the strongest binding sites, prior to reaction with atomic H, so that the chemically-produced water molecules would automatically be formed in strong binding sites that would otherwise correspond to nascent surface inhomogeneities, thus allowing the ice to form smoothly at microscopic scales. Whether the production of compact water ice is due to the thermal diffusion of atomic oxygen, or to non-thermal, reaction-induced diffusion of the water molecule itself, the chemical production of water on dust grains appears to lead to little or no porous structure, in agreement with the existing observational evidence \citep[although it has also been suggested that the filling of water-ice pores with molecules such as CO could remove the dangling-bond feature; e.g.][]{He19}.

Similarly, until recently there has been little direct evidence to indicate either the degree or the precise form of porosity, if any, that interstellar dust-grains themselves might assume; however, some indirect evidence suggests that interstellar dust grains could be highly porous. While models of scattering by dust typically assume optical properties corresponding to spherical or spheroidal dust grains of uniform composition, \citet{mathis96} tested models using composite dust grains, whose volume could also be composed of some fraction of empty space, $f_{vac}$. Mathis found that models using grains with $f_{vac} = 0.45$ could be successful in reproducing observational extinction curves (although see also Weingartner \& Draine 2001); this vacuum fraction, which we identify with the porous structures that we model here (Sec. \ref{results}), was assumed in those models to be present equally in grains of all sizes.

Dust grains detected in the comae of comets may also be porous, based on analysis of the range of grain sizes and densities. In the case of comet 67P, these grains appear to be loose aggregations of low density (0.1 -- 1 g cm$^{-3}$) and mean size around 10$\mu$m that are made up of smaller sub-units \citep{agarwal}. The sub-units themselves are of sizes more appropriate to interstellar grains, on average around 0.1$\mu$m, and are thought to be closer in density to the native material from which they are composed \citep{mannel}, indicating a low degree of porosity on those size scales. It is unclear, however, whether those sub-units could themselves be formed from smaller structures.

Dust-grain coagulation models appropriate to star- and planet-forming environments \citep{ossenkopf,ormel} indicate that large, micron-scale grains can become porous, but their smallest sub-units are of similar size to the typical interstellar dust grains, and therefore do not indicate whether any porosity may exist in those smaller grains. Other dust models \citep{jones} have considered the accretion of carbon atoms onto existing grains in the outer regions of molecular clouds, as a mechanism to replenish the dust destroyed through energetic processes such as interstellar shocks, in order to explain the apparent imbalance between dust formation and destruction in the ISM. This accretion process, which would necessarily occur at low dust temperatures (no greater than around 30~K under typical interstellar conditions), could plausibly introduce irregularities to the grain surfaces, or, with a sufficient degree of accretion, outright porosity in the outer layers of the grain material.

Most recently, \citet{potapov} carried out experiments in which interstellar dust-grain analog particles composed of amorphous carbon were formed and deposited onto a substrate. Ices of water, CO, and other molecular species were then deposited onto the grains, and their temperature-programmed desorption profiles analyzed. The authors reported behavior indicating that the ices were spread over very large surface areas, indicating an extreme degree of porosity. They proposed that interstellar grains may have surface areas enhanced by factors of several hundred, producing maximal ice thicknesses on the order of 1 monolayer in interstellar clouds. However, we note that the degree of porosity in interstellar grains themselves would be dependent on the precise mechanism and environment of their formation.

Here, we investigate the influence of interstellar dust porosity on the chemistry and composition of ices that form on the dust-grain surfaces. Dust porosity has the potential to affect the chemistry occurring on dust-grain surfaces, via increased surface area, enhanced local binding energies, and the possibility of trapping of molecules within the pores as ice mantles build up on the grains.

\citet{garrod13a}, mentioned above, introduced a microscopic Monte Carlo chemical kinetic model for interstellar grain-surface chemistry that was capable of simulating diffusive chemistry on fully three-dimensional surfaces. The model uses an off-lattice treatment, allowing surface particles to assume positions that are determined entirely by their interactions with nearby binding partners. In that study, the model was used to simulate the production of water and several associated molecules over astronomical timescales; an important effect noted by Garrod was the formation of porosity in the ices, which was dependent on both the temperature of the dust grain and the rate of accretion of gas-phase atoms onto the grain, as controlled by the gas density; under the dark-cloud conditions used in those models, the ices produced were found to be non-porous. Here, we use an updated version of the model with a larger chemical network, such that all of the main interstellar ice components may be modeled. The structure of the underlying dust grain may be directly controlled by the placement of the (carbon) atoms from which it is constructed, meaning that any structure can in principle be used; this includes the possible variation of both the large-scale morphology of a grain and its microscopic surface roughness. We run chemical models on both porous and non-porous grains of a selection of sizes, using broadly spherical large-scale structures. 
These models represent the first such attempt to model explicitly the effects of dust-grain porosity on interstellar grain-surface chemistry.

\section{Methods}

The model results presented here make use of the off-lattice microscopic Monte Carlo chemical kinetics code {\em MIMICK} (Model For Interstellar Monte Carlo Ice Chemical Kinetics), developed by \citet{garrod13a}. This model uses an exact description of a dust grain in which the positions of all of its constituent atoms are explicitly described. The model traces the chemistry occurring on the dust-grain surface by similarly tracking the positions and movements of the atoms and molecules that are adsorbed onto the surface from the gas phase. The positions of those species on the surface correspond to local potential minima (binding sites), as determined by the strength of the pairwise potentials experienced by each surface particle. Positions may change when mobile species make thermal hops between adjacent binding sites. 

All of the various physical and chemical processes occurring on the grain, namely adsorption/accretion, thermal desorption, thermal hopping and reaction, are treated as a sequence of consecutive events. The Monte Carlo treatment used here follows that of \citet{gillespie}, in which random numbers are used to determine (i) which process occurs next and (ii) the time increment between each such event, based on the calculated rates of every possible process at that moment in time. In the case of thermal diffusion, each individual hop from one binding site to another is treated as a distinct event. Standard rate formulations are used for each of the various processes for each individual atom/molecule on the surface; thus, those rates are dependent not only on the type of atom/molecule in question but on the number, type, and spatial arrangement of its binding partners. 

The model employed here uses a somewhat more advanced treatment for the calculation of surface binding energies and diffusion barriers than that of \citet{garrod13a}. \citet{clements} provide an outline of this treatment, which is described in detail by Garrod \& Deselm ({\em in prep.}). In brief, each pairwise interaction between atoms, molecules or surface particles is treated as a Lennard-Jones (6--12) van der Waals potential, with a defined optimal separation, $\sigma$, and a potential well depth, $\epsilon$. When an atom or molecule arrives on the surface through adsorption from the gas phase, its position is optimized by minimization of the sum of all pairwise potentials that it experiences (while the positions of the individual binding partners are assumed to be fixed). The 6--12 potentials are tuned to have a cutoff separation of $r=2.5\sigma$; any nearby species that fall within this cutoff distance are termed ``binding partners'', and thus contribute to the total binding energy of the adsorbed species. Any binding partners that fall within a distance $r=1.5\sigma$ are termed ``contiguous'', and may be considered to be in direct contact with the species in question.

The process of diffusion may occur in multiple possible directions for a given surface atom/molecule, with each direction having its own associated energy barrier. Firstly, the model determines which directions are possible. To do this, all viable diffusion pathways are assumed to pass through saddle-points in the local surface potential; due to the spherical symmetry of the pairwise potentials, each saddle-point falls approximately between adjacent pairs of binding partners that are immediately contiguous to the diffuser \citep[see][for a diagram of such an arrangement]{garrod13a}. Diffusion through that saddle-point may be imagined as a rotation of the diffuser about the axis that joins two of those contiguous binding partners to each other; those binding partners are termed the ``saddle-point pair''. Each saddle-point pair can be identified geometrically, and has a unique diffusion pathway associated with it. \citep[][also provides a more detailed description of the geometric considerations involved in identifying the saddle-point pairs]{garrod13a}. As an example, one may imagine an atom sited atop a triangular arrangement of three contiguous binding partners. Any given pair of those binding partners would constitute a saddle-point pair, with rotation around each such pair providing a unique direction of diffusion through a potential saddle point.

For every atom or molecule on the surface at a given moment, the model holds a list of saddle-point pairs (i.e. diffusion directions), for each of which a diffusion barrier is also defined. A diffusion barrier consists of the energy required to break or stretch each surface bond to the furthest distance the diffuser could reach as a result of rotation around the axis defined by the saddle-point pair. Some existing bonds would be entirely broken by such a procedure, while others would merely be stretched, while still remaining within the cut-off distance of the van der Waals interaction. The calculation excludes any bonds that would be shortened in this procedure. The bonds between the diffuser and the binding partners that comprise the saddle-point pair also do not contribute to the diffusion barrier, as the rotation around the saddle-point pair occurs at constant radial distance from each. The diffusion barriers are therefore always lower than the total binding energy, and they vary depending on the direction of diffusion; diffusion events that require the breaking or stretching of stronger pairwise potentials have higher barriers and are therefore less likely to occur. \citep[The key difference between the present treatment and that of][is that the older model only considered contiguous binding partners as contributors to the diffusion barriers and binding energies]{garrod13a}. If and when a particular diffusion event is selected by the Monte Carlo routine according to its barrier-defined rate, the diffusing particle is allowed to rotate around the saddle-point pair until it encounters another particle on the other side; its position is then re-optimized based on all of the pairwise potentials experienced in its new location. 

The full diffusion procedure described above, while being approximate in the determination of diffusion barriers and prescriptive in the choice of available directions, obviates the need to evaluate directly the potential energy surface involved in every possible diffusion event of every surface species; such a simplification is essential to allow this kinetic model to run on manageable timescales.

All surface diffusion considered in this model is based on thermal rates (as defined by the barriers described above); no tunneling contributions to diffusion rates are considered. At the 10~K temperature used in the models presented here, tunneling is not expected to contribute significantly to the diffusion rates of hydrogen atoms; quantum calculations by \citet{Sen17} indicate a crossing temperature (i.e. the temperature at which tunneling begins to dominate) of $T_C($H$)=9.7$~K for H atoms on amorphous water ice. $T_C$ values for other species that could plausibly diffuse via tunneling, e.g. H$_2$, should be expected to be substantially lower, due to the larger mass involved; the same authors found $T_{C}($D$)=6.9$~K on amorphous water ice.

In the present version of the {\em MIMICK} model, all surface particles, whether atoms or molecules (or atoms comprising the dust grain itself) are considered to be spherical, with a radius of 1.6~\AA, such that the optimal separation used in the pairwise potentials is $\sigma=3.2$~\AA. This size is chosen to ensure that the physical size of the ice formed on the grains is representative of water ice, which is the dominant constituent. Variable values of $\sigma$ will be introduced in future work.

Importantly, as in the models of \citet{clements}, the thermal desorption of a surface particle follows a specific trajectory, which is defined by the vector sum of the pairwise potentials at the starting position. The desorbing particle incrementally traces out this trajectory until it reaches the edge of the calculation-space used in the model. If the trajectory intersects with other particles then the desorbing particle may be re-accreted. This eventuality is especially important in cases where enclosed surfaces exist within the porous structures of the grain. Note that the present model does not use the treatment presented by Clements et al. for the gradual relaxation of a newly-adsorbed particle. Atoms and molecules accreted from the gas phase are assumed to thermalize immediately on the surface.

Only surface particles that are considered ``free'' are allowed to diffuse or to desorb. Free particles are defined as those that have available at least one surface diffusion pathway, as determined by their local geometry. Particles that are completely enclosed by contiguous binding partners (as would be the case for molecules within the bulk ice) have no such pathways. Thus, any particle that is capable of diffusion also has some probability of thermal desorption. Counting the number of free particles also provides a determination of the number of particles present in the surface layer of the ice.

Reactions between surface species are allowed to occur when the reactants become contiguous, whether that be through diffusion, adsorption, or as the result of an earlier reaction. Once contiguous, that reaction between specific particles is considered as a process that competes with all others to be the next event in the sequence. If the reaction has no activation energy barrier, then it is assigned a rate equal to the sum of the characteristic vibrational frequencies of the reactants. The rates of reactions that have activation energy barriers are further moderated by an efficiency factor representing the faster of either the thermal reaction probability (Boltzmann factor) or a probability based on tunneling through a rectangular barrier, as is typically employed in the usual rate-equation based models \citep[e.g.][]{hasegawa}.

Many typical (non-MC) kinetic models of grain-surface chemistry include a chemical desorption (CD) mechanism, whereby the energy released in the formation of a molecule via a surface reaction is sufficient to allow the molecule--surface bond to be broken and the reaction product to desorb into the gas phase \citep{Garrod07}. This mechanism is usually assumed to occur with an efficiency of approximately (or sometimes precisely) 1\%; in the remainder of cases, the reaction energy would be lost through vibrational interactions with the surface and the product would remain on the grain. Experimental studies \citep{Minissale16} have confirmed that this process can occur, typically with greater probability on bare-grain analogs than on the amorphous water ice that is expected to build up on the grain surfaces. For most reactions, those experiments obtained only upper limits to the efficiency, on the order of 10\%. Recent {\em ab initio} molecular dynamics studies of the H + CO $\rightarrow$ HCO reaction system on crystalline water ice indicate that energy loss to the surface occurs too rapidly to allow product desorption \citep{Pantaleone20}. In most kinetic models that include the CD mechanism, while its inclusion may be important to the abundances of certain gas-phase species (such as methanol) that do not apparently have a viable gas-phase formation mechanism, the influence on the abundances of surface species is relatively minor. We therefore leave the inclusion of this desorption mechanism in the {\em MIMICK} model to future work.
\\

\subsection{Construction of the dust grains}

Three sizes of non-porous grain are constructed for use in the chemical model, with radii of approximately 50, 75, and 100~\AA. In each case, this begins with the generation of a spherical grain composed of atoms in a simple cubic arrangement, placed at their optimal separations, $\sigma$. Because the chemical model {\em MIMICK} requires only the surface of the grain to be defined, a sphere of material is removed from the center, leaving a shell that is substantially thicker than the cutoff distance of the potentials used by {\em MIMICK}. The process up to this point is identical to that used by \citet{garrod13a}, albeit with much larger grains in the present case. 

To produce an amorphous surface structure, the simple-cubic grain is first compressed in all three dimensions by a factor 0.75. The position of each atom is then offset in a randomized direction, by a random distance ranging from 0 -- $\sigma$. These two procedures randomize the positions of the atoms sufficiently to remove all trace of the original simple cubic micro-structure, while retaining the spherical shape. Each atom in the grain is then allowed to relax, according to a randomized sequence, using Lennard-Jones type potentials of arbitrary, uniform strength. This ensures that all of the atoms in the grain are at appropriate distances from each other. The upper row of Fig. 1 shows the non-porous grains produced in this way; the surface of each is amorphous, while retaining a well defined spherical morphology at large scales.

To produce a porous grain, a smaller grain core is initially created using the method described above.
The largest porous grain uses the smallest non-porous grain as its core, while the medium-sized porous grain uses a core of radius 25~\AA. The chemical model is then employed to deposit more (carbon) atoms from the gas phase onto this core surface. The trajectories of the incoming atoms are entirely randomized. Upon landing, no chemistry or surface diffusion is allowed, but the positions of the accreting atoms are optimized, again using Lennard-Jones potentials. This hit-and-stick type approach results in open, porous structures. The accretion of atoms is halted when the maximum radial position of any atom (centered on the zero-position of the grain core) is equal to the maximum radial position of atoms in either the intermediate- or the large-sized non-porous grain; the resulting porous portions of each grain have a maximum thickness $\sim$50~\AA, but the average is closer to 40~\AA. The porosity of the porous portions of both grains is measured to be $\sim$64~\%. Taking into account the non-porous core of each grain, this produces overall porosities of 61~\% and 53~\% for the medium and large grains, respectively. The latter is not far from the value used by Mathis (1996).

\begin{figure}[ht]
    \centering
    \includegraphics[width=0.6\textwidth]{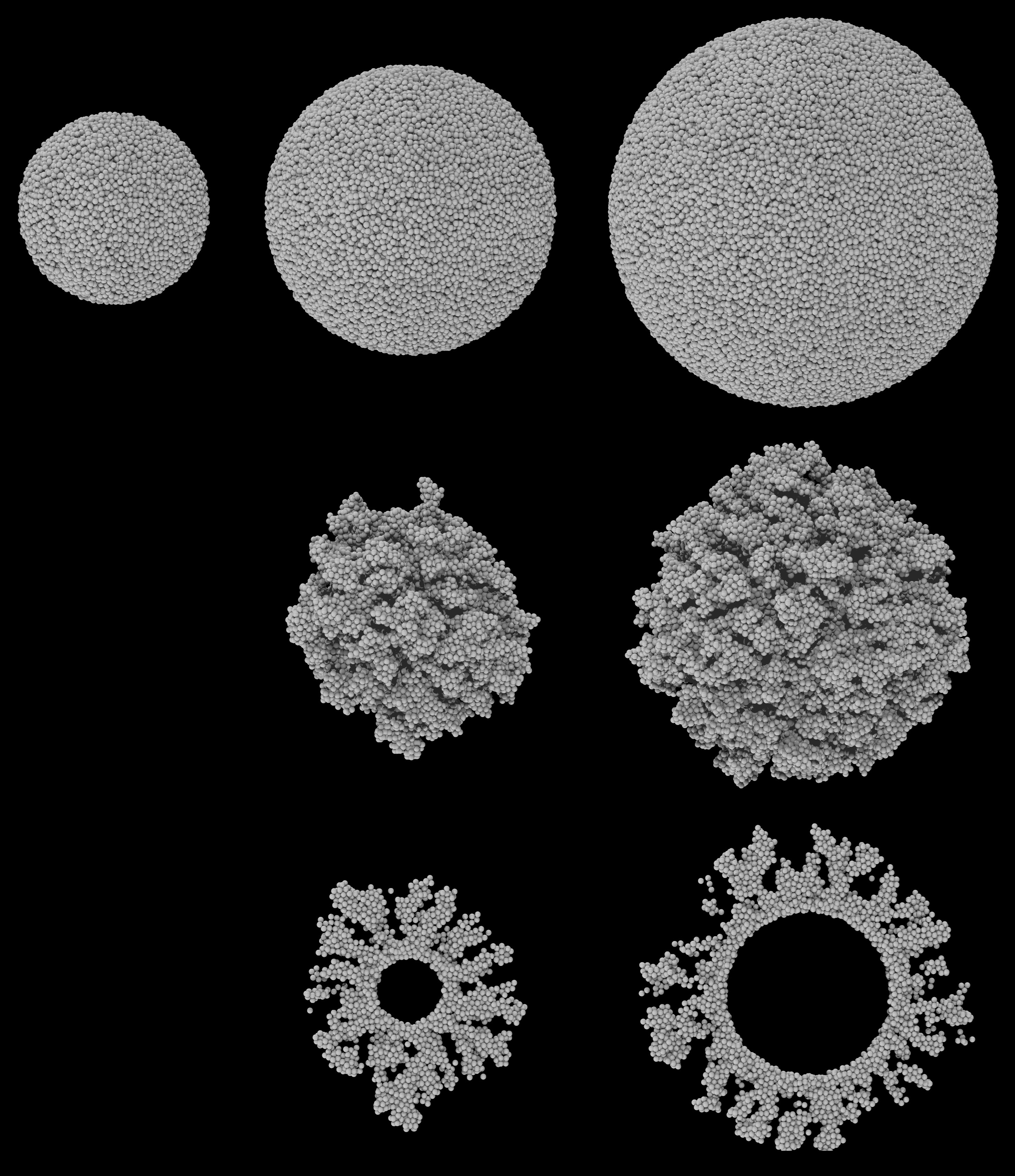}
    \caption{Ray-traced images of the five grain types used in the chemical models. All grains are composed of a collection of spheres representing carbon atoms. Top row: the three non-porous grains; middle row: the two porous grains, of comparable size to the two largest non-porous grains; bottom row: the cross sections of the two porous grains. From left to right, grains have a radius approximately 50, 75, and 100 \AA. Note that the inner portions of the grains are removed to simplify the chemical kinetics calculations; the inner void is completely inaccessible to surface atoms and molecules.}
    \label{figgrains}
\end{figure}

The five different grains created for the chemical simulations range in size from 50 -- 100~\AA\, in radius. This is smaller than the typical value of 0.1~$\mu$m (1000~\AA) used in chemical models, although smaller grains such as these would be more numerous in the distribution and should provide a larger fraction of the available surface area for accretion, assuming non-porous, spherical grains. Smaller grains are also more computationally favorable using the microscopic Monte Carlo kinetics approach, as larger grains generate larger ices composed of greater numbers of atoms and molecules, each of which must be handled individually. 

During the main chemical simulations, the atoms comprising the dust grains are fixed in place and do not participate in any chemical reactions, either directly or as a catalyst. Their main function is to define the physical surface of the grain.
\\

\subsection{Physical and chemical conditions}

To represent a dark interstellar cloud (or, more precisely, a dark cloud core), the gas density is set to $n_{\mathrm{H}} = 2 \times 10^4$ cm$^{-3}$ and both the grain and gas temperatures are set to 10~K \citep[e.g.][]{BerginTafalla07}; similar conditions have been used in past models of dark cloud chemistry \citep[e.g.][]{hasegawa,Garrod07}. Although no gas-phase chemistry is included in these models, the gas density and temperature determine the rates of accretion of gas-phase species onto the grain.

The gas-to-dust ratio by number is tailored for the specific grain size used in each model run, to maintain a consistent mass ratio of 100:1 appropriate to the Milky Way. This influences the total number of atoms initially available in the gas phase in our models. The chosen grain size used in each model run is assumed to be representative of the entire dust population, assuming a mass density of 3~g~cm$^{-3}$ for grain material. As grain size increases, the number of representative dust grains must decrease, thus increasing the amount of gas per dust grain. For the large grain, a gas-to-dust ratio by number of $10^{10}$ is used, which corresponds specifically to the number of hydrogen atoms present in either H or H$_2$ at the start of the models. The medium and small grains use lower values, as scaled by volume. For comparability between the chemical models, the gas-to-dust ratios used for the large- and medium-sized porous grains are the same as those of their non-porous equivalents.

The total budget (i.e. number of atoms) of each element used in the model is generated based on the gas-to-dust ratio and the abundance of each, relative to total hydrogen. The assumed gas-phase composition at the beginning of each model is shown in Table \ref{tab:initial}, based on the initial gas-phase elemental abundances used by \citet{garrod13b}. The grains in our models begin with no ice mantles. To better represent the composition of ices in dark clouds, a proportion of the carbon and oxygen budgets is incorporated into CO. Most of the hydrogen budget is assumed to be in the form of molecular hydrogen, with the atomic H abundance also chosen to represent typical dark cloud values. Gas-phase H$_2$ is not allowed to adsorb onto the grains, following \citet{garrod13a}; this simplification is necessary to allow the models to run on manageable timescales.

As the models progress, atoms and molecules in the gas accrete onto the grain surface, depleting the gas-phase budget of each and thus reducing the amount still available for the further growth of the ice mantles. However, an exception is made in the treatment of the atomic H budget. To ensure that its rate of adsorption is always appropriate to dark cloud values, its abundance in the gas phase is held steady at its initial value. In full gas-grain models of dark cloud chemistry, gas-phase processes such as cosmic-ray induced dissociation allow the gradual release of some hydrogen locked up in H$_2$ back into atomic form, meaning that the fractional abundance of H is maintained at some base value, even though other species may become strongly depleted from the gas phase over time. This simple fix to the code avoids having explicitly to model the gas-phase chemistry.

\begin{table}[]
    \centering
    \caption{Initial gas-phase abundances as a fraction of total hydrogen.}
    \begin{tabular}{l|l}
         Chemical species & Fractional abundance \\
\hline
         H$_2$ & $0.4999$ \\
         H & $2.0 \times 10^{-4}$ (fixed) \\
         O & $2.7 \times 10^{-4}$ \\
         C & $9.0 \times 10^{-5}$ \\
         N & $7.5 \times 10^{-5}$ \\
         CO & $5.0 \times 10^{-5}$ \\
    \end{tabular}
    \label{tab:initial}
\end{table}

The model assumes that the sticking coefficient for all species is unity, following typical assumptions used in other models. Experimental work by \cite{He16b} indeed found a sticking coefficent of 1 for CO (the only adsorbing molecule in our model), although their low-temperature value for H$_2$ was in the 0.6--0.7 range. The sticking coefficient for atomic H on an amorphous water surface was calculated to be 0.980 by \citet{Dupuy16} at dust/gas temperatures of 10~K. Sticking coefficients for H on amorphous carbon surfaces are not well defined, although \citet{Pirronello97} suggest it could be appreciably less than one. Experiments by \citet{Chaabouni12} find a value of 1 for silicate surfaces, however.

The model includes 39 chemical species and 70 different reactions, allowing C, O, N and CO to be hydrogenated all the way to CH$_4$, H$_2$O, NH$_3$ and CH$_3$OH. Molecules such as H$_2$, C$_2$, O$_2$, N$_2$ and CO may also be formed on the grain surfaces, along with various others. Table \ref{tab:network} details each of the reactions included in the model, with the assumed activation energy barriers and barrier widths indicated; this network is a simplified version of that presented by \citet{garrod13b}. Most of the reactions occurring on the grains are initiated by the diffusion of atomic H to meet its reaction partners, although reactions may occur in any situation in which two reactants come into contact. Table \ref{tab:potentials} provides a selection of important pair-wise potentials used in the model, which determine both the binding energies and diffusion barriers experienced locally by surface species. The potentials are somewhat lower than those used by \citet{garrod13a}, which reflects the fact that contributions to the binding of all local species within the cutoff distance are now considered, not only those species that are contiguous to the particle in question, which was the previous method. As described by \citet{garrod13a}, the pairwise potentials are intended to reproduce the overall binding energies of the species in question on surfaces of, for example, pure amorphous water.

\begin{table}[]
    \centering
    \caption{The reaction network used in MIMICK. Activation barriers, where present, are in units of Kelvin, with barrier widths in Angstrom.}
    \resizebox{\columnwidth}{!}{%
    \begin{tabular}{ l | c | c || l | c | c }
            Reaction &  E$_a$ (K) & Width (\AA) &  Reaction &  E$_a$ (K) & Width (\AA) \\
\hline
        H + H ${\rightarrow}$ H$_2$ &  - &  - &  O + CH ${\rightarrow}$ HCO &  - &  -  \\
        H + O ${\rightarrow}$ OH &  - &  - &  O + CH$_2$ ${\rightarrow}$ H$_2$CO &  - &  -  \\
        H + OH ${\rightarrow}$ H$_2$O &  - &  - &  O + CH$_3$ ${\rightarrow}$ CH$_3$O &  - &  -  \\
        H + O$_2$ ${\rightarrow}$ O$_2$H &  - &  - &  O + CO ${\rightarrow}$ CO$_2$ &  1000 &  1.27  \\
        H + O$_2$H ${\rightarrow}$ H$_2$O$_2$ &  - &  - &  O + HCO ${\rightarrow}$ H + CO$_2$ &  - &  -  \\
        H + O$_3$ ${\rightarrow}$ OH + O$_2$ &  - &  - &  OH + OH ${\rightarrow}$ H$_2$O$_2$ &  - &  -  \\
        H + H$_2$O$_2$ ${\rightarrow}$ OH + H$_2$O &  1800 &  1.00 &  OH + CH$_2$ ${\rightarrow}$ CH$_2$OH &  - &  -  \\
        H + C ${\rightarrow}$ CH &  - &  - &  OH + CH$_3$ ${\rightarrow}$ CH$_3$OH &  - &  -  \\
        H + CH ${\rightarrow}$ CH$_2$ &  - &  - &  OH + CH$_3$OH ${\rightarrow}$ H$_2$O + CH$_2$OH &  359 &  1.00  \\
        H + CH$_2$ ${\rightarrow}$ CH$_3$ &  - &  - &  OH + CH$_3$OH ${\rightarrow}$ H$_2$O + CH$_3$O &  852 &  1.00  \\
        H + CH$_3$ ${\rightarrow}$ CH$_4$ &  - &  - &  OH + CH$_4$ ${\rightarrow}$ H$_2$O + CH$_3$ &  1780 &  1.00  \\
        H + N ${\rightarrow}$ NH &  - &  - &  OH + CO ${\rightarrow}$ H + CO$_2$ &  80 &  1.00  \\
        H + NH ${\rightarrow}$ NH$_2$ &  - &  - &  N + N ${\rightarrow}$ N$_2$ &  - &  -  \\
        H + NH$_2$ ${\rightarrow}$ NH$_3$ &  - &  - &  CH$_3$ + CH$_3$ ${\rightarrow}$ C$_2$H$_6$ &  - &  -  \\
        H + CO ${\rightarrow}$ HCO &  2320 &  1.35 &  CH$_3$ + HCO ${\rightarrow}$ CH$_3$CHO &  - &  -  \\
        H + HCO ${\rightarrow}$ H$_2$ + CO &  - &  - &  HCO + HCO ${\rightarrow}$ HCOCHO &  - &  -  \\
        H + HCO ${\rightarrow}$ H$_2$CO &  - &  - &  CH$_3$ + CH$_3$O ${\rightarrow}$ CH$_3$OCH$_3$ &  - &  -  \\
        H + H$_2$CO ${\rightarrow}$ CH$_2$OH &  4500 &  1.35 &  CH$_3$ + CH$_2$OH ${\rightarrow}$ C$_2$H$_5$OH &  - &  -  \\
        H + H$_2$CO ${\rightarrow}$ CH$_3$O &  2320 &  1.35 &  HCO + CH$_3$O ${\rightarrow}$ HCOOCH$_3$ &  - &  -  \\
        H + H$_2$CO ${\rightarrow}$ H$_2$ + HCO &  2960 &  1.22 &  HCO + CH$_2$OH ${\rightarrow}$ HCOCH$_2$OH &  - &  -  \\
        H + CH$_2$OH ${\rightarrow}$ CH$_3$OH &  - &  - &  C + C ${\rightarrow}$ C$_2$ &  - &  -  \\
        H + CH$_2$OH ${\rightarrow}$ H$_2$ + H$_2$CO &  - &  - &  C + CH ${\rightarrow}$ C$_2$H &  - &  -  \\
        H + CH$_3$O ${\rightarrow}$ CH$_3$OH &  - &  - &  C + CH$_2$ ${\rightarrow}$ C$_2$H$_2$ &  - &  -  \\
        H + CH$_3$O ${\rightarrow}$ H$_2$ + H$_2$CO &  - &  - &  C + CH$_3$ ${\rightarrow}$ C$_2$H$_3$ &  - &  -  \\
        H + CH$_4$ ${\rightarrow}$ H$_2$ + CH$_3$ &  5940 &  2.17 &  CH + CH ${\rightarrow}$ C$_2$H$_2$ &  - &  -  \\
        H$_2$ + C ${\rightarrow}$ CH$_2$ &  2500 &  1.00 &  CH + CH$_2$ ${\rightarrow}$ C$_2$H$_3$ &  - &  -  \\
        H$_2$ + CH$_2$ ${\rightarrow}$ CH$_3$ + H &  3530 &  1.00 &  CH + CH$_3$ ${\rightarrow}$ C$_2$H$_4$ &  - &  -  \\
        H$_2$ + CH$_3$ ${\rightarrow}$ H + CH$_4$ &  6440 &  1.00 &  CH$_2$ + CH$_2$ ${\rightarrow}$ C$_2$H$_4$ &  - &  -  \\
        H$_2$ + NH$_2$ ${\rightarrow}$ H + NH$_3$ &  6300 &  1.00 &  CH$_2$ + CH$_3$ ${\rightarrow}$ C$_2$H$_5$ &  - &  -  \\
        H$_2$ + OH ${\rightarrow}$ H + H$_2$O &  2100 &  1.00 &  H + C$_2$ ${\rightarrow}$ C$_2$H &  - &  -  \\
        O + O ${\rightarrow}$ O$_2$ &  - &  - &  H + C$_2$H ${\rightarrow}$ C$_2$H$_2$ &  - &  -  \\
        O + OH ${\rightarrow}$ O$_2$H &  - &  - &  H + C$_2$H$_2$ ${\rightarrow}$ C$_2$H$_3$ &  1300 &  1.00  \\
        O + O$_2$ ${\rightarrow}$ O$_3$ &  - &  - &  H + C$_2$H$_3$ ${\rightarrow}$ C$_2$H$_4$ &  - &  -  \\
        O + O$_2$H ${\rightarrow}$ O$_2$ + OH &  - &  - &  H + C$_2$H$_4$ ${\rightarrow}$ C$_2$H$_5$ &  605 &  1.00  \\
        O + C ${\rightarrow}$ CO &  - &  - &  H + C$_2$H$_5$ ${\rightarrow}$ C$_2$H$_6$ &  - &  -  \\
    \end{tabular}%
    }
    \label{tab:network}
\end{table}

The barriers against diffusion in any direction are always lower than the corresponding binding energies for a particular particle in a particular binding site, although different directions of diffusion typically have different values. The models produce as an output the distribution of the ratios of diffusion barrier versus binding energy for each chemical species in the network (in increments of 0.05), which are calculated at every output time in the model. In around 99\% of atomic-H surface diffusion events, the barrier of the successful diffusion direction lies in the range 0.25--0.7 times the local binding energy, with around 18-20\% of those barriers falling in the 0.4-0.45 range, which is the most populous bin. The average binding energy of an H atom on the final (outer) surface of the ice mantles is approximately 460~K. This compares well with the value used in many past gas-grain models \citep[e.g.][]{garrod13b} of 450~K; DFT calculations by \citet{Wakelam17b} indicate values of 400~K and 680~K for H on water, dependent on the method used. Using our model's average binding energy of 460~K, the diffusion barriers in the above fractional range produced by the model provide absolute values in the range 115--322~K, with the most common falling in the 184--207~K bin. \citet{Sen17}, based on their calculations for H on amorphous water ice, recommended a representative ratio of diffusion barrier to binding energy for H of 0.37; combined with their representative binding energy of 661~K (57~meV), this suggests a diffusion barrier of 243~K. This latter value in particular seems fairly consistent with our range of values, in spite of the higher overall binding energy. Indeed, looking at the actual distribution of diffusion barriers produced by those authors, their most common diffusion barriers lie in the $\sim$175 -- 230~K range, which coincides well with our most common bin.

For H$_2$, the model produces average binding energies of around 380--395~K; \citet{garrod13b} use a value of 430~K. The calculations of \citet{Wakelam17b} for the H$_2$--H$_2$O dimer suggest a binding energy of 800~K, but full cluster calculations by \citet{Ferrero20} for amorphous solid water give a value in the range 226--431~K, in keeping with our model values. The experimental measurements of \citet{He16a} indicate binding energies in a range between 322 and 505~K, with the former corresponding to high H$_2$ coverage and the latter to low coverage. Our models produce binding energies for CO and O$_2$ of around 1015~K and 1240~K, respectively; these are somewhat higher and somewhat lower, respectively, than the calculations by \citet{Ferrero20} although our values fall comfortably within the range of experimental values quoted by those authors (870--1600~K for CO and 914--1520~K for O$_2$).

For H$_2$O, the model produces an average binding energy around 4,750~K. The outer ice surface contains a selection of molecules besides water, so the effective binding energy of a water molecule is somewhat lower than might be expected on a pure water ice; 
a common value used in rate-equation models is 5,700~K \citep{garrod13b}. The calculations of \citet{Wakelam17a} indicate values of 4800 or 5300~K based on water dimers or clusters, while \citet{Ferrero20} obtain a range of 3605--6111~K. Experiment also provides a range of values, although \cite{Sandford90} obtain 4815~K for unannealed water ice that is very close to our value. Again, we note that our value is produced with a water-dominated ice mixture that includes all of the typically observed interstellar solid-phase species, rather than pure water.
In practice, at the low temperatures used in the present models, the rates of diffusion and desorption of water molecules are negligible.

The model does not consider the influence of the condensation energies of adsorbed species, the energy released by chemical reactions, nor the effects of high-energy photons on the overall temperature of the grain/ice. Although the associated temperature fluctuations could be substantial for very small grains \citep[e.g.][]{cuppen06}, they are beyond the scope of the present model. The recent modeling simulation of water-ice deposition by \citet{clements} included a treatment for the non-thermal diffusion of newly-adsorbed molecules, caused by the excess energy gained as they are accelerated into the surface potential. That treatment was tuned specifically for water molecules, and we do not extend it here to other species. In any case, the ability of the adsorbed atoms in the present model to exhibit some degree of thermal diffusion following adsorption makes this simplification less important under the conditions tested here.

\begin{table}[]
    \centering
    \caption{Pairwise potentials, $\epsilon$ (K), between selected surface species.}
    \begin{tabular}{ l | r | r | r }
                 &  Grain carbon & H$_2$O & CO \\
\hline
        H        &   45 &  45 &  45  \\
        H$_2$    &   40 &  40 &  40  \\
        C        &  100 & 100 & 100  \\
        O        &  100 & 100 & 100  \\
        N        &  100 & 100 & 100  \\
        OH       &  300 & 450 & 120  \\
        CO       &  120 & 120 & 120  \\
        N$_2$    &  125 & 125 & 120  \\
        O$_2$    &  150 & 150 & 120  \\
        CO$_2$   &  280 & 280 & 120  \\
        CH$_4$   &  180 & 180 & 120  \\
        NH$_3$   &  520 & 700 & 120  \\
        CH$_3$OH &  320 & 640 & 120  \\
        H$_2$O   &  350 & 700 & 120  \\
    \end{tabular}
    \label{tab:potentials}
\end{table}

The models are run for simulated times on the order of tens of thousands of years, and by the end of each model run around half of the heavy atoms have been depleted from the gas phase, forming ices on the grain surfaces. By the end of its run-time, each model is thus on the cusp of experiencing strong gas-phase depletion which would slow the further growth of the ice mantle. The time taken to build up an ice mantle of a particular size is dependent on the cross-sectional area presented by the combined grain and ice mantle, averaged over all solid angles \citep[the details of the accretion process are described in detail by][]{garrod13a}. The different sizes and morphologies of the grains thus affect the cross-sectional area presented to gas phase species, while those cross sections also rise as the ice mantles grow over time. For purposes of comparison between models, it is therefore helpful to define output times in relative terms, corresponding to the amount of depletion from the gas phase. This measure is more instructive than, for example, a determination of the number of ice molecules on the grains or the thickness of the ice, as it makes no assumptions about the chemical or physical form that the ice mantles take. Each model is set up to produce outputs for 1000 different time values, which here we will refer to as ``intervals''. Those intervals are evenly spaced in the amount of O, C, N and CO that have been accreted from the gas phase as a fraction of the initial amount (which varies depending on the size of the grain, via the gas-to-dust ratio). Each interval corresponds to a roughly equivalent amount of ice build-up, but as time goes on and depletion from the gas phase increases, the timescale required to complete one interval lengthens. The final populations of the grain-surface molecules achieved in each grain-size model scale roughly with the volume ratio of the underlying grains.

The run-time (wall time) of the models ranges from a few days to two weeks, depending on the size of the dust grain.

For each grain size and morphology used in the models, a total of three physically and chemically identical runs are carried out, with the only difference being the initial random number seed used in the Monte Carlo algorithm. This ensures that the model results are reproducible, and to some extent demonstrates the degree of random variation in the chemical abundances.

\begin{figure}[ht]
    \centering
    \includegraphics[width=0.99\textwidth]{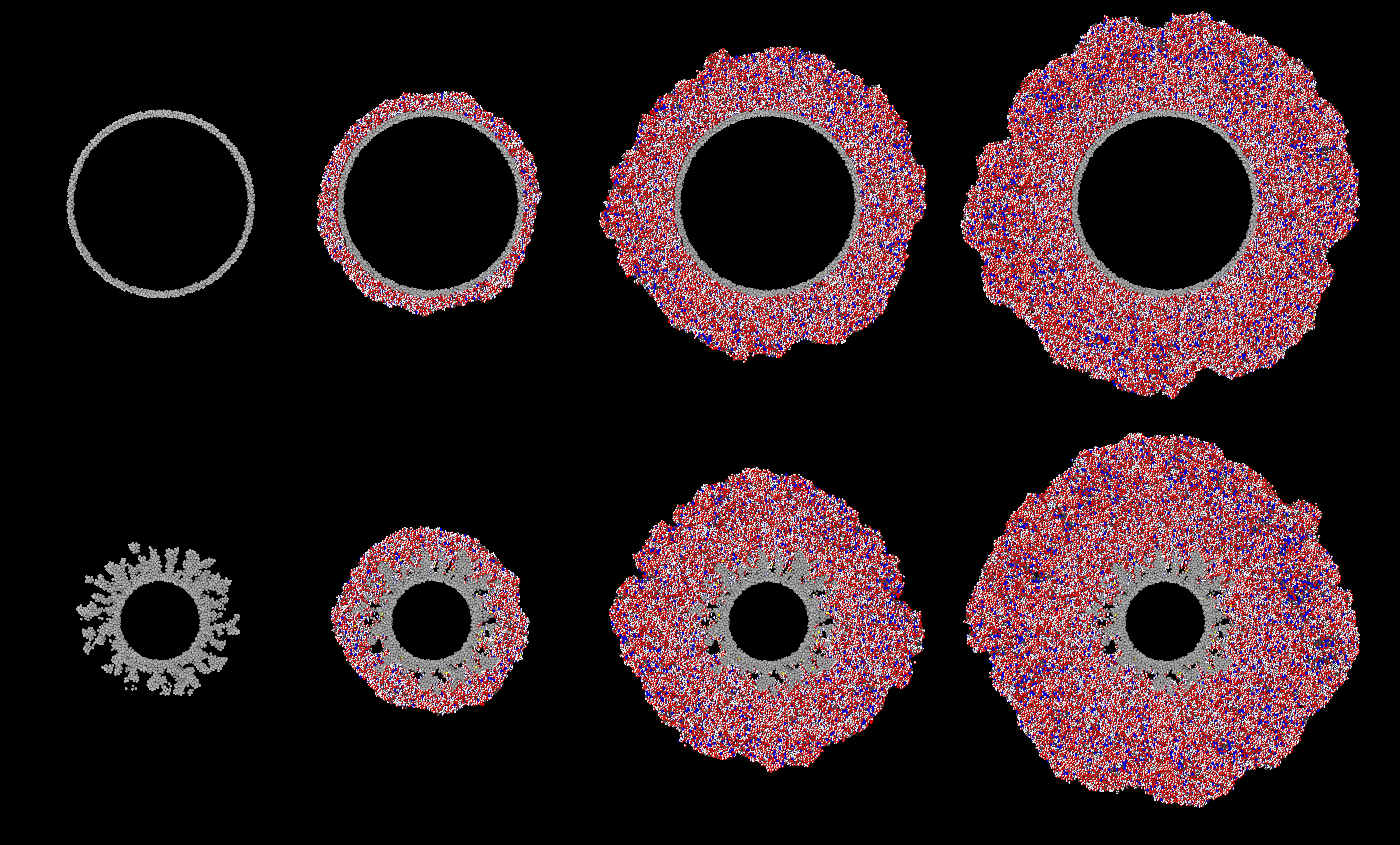}
    \caption{Ray-traced, cross-sectional images of the large (100 {\AA} radius) non-porous (top) and large porous (bottom) grain types at different stages of ice build-up; left to right: interval 0, interval 100, interval 500, interval 1000. Gray spheres represent the carbon atoms composing the dust grain. Grain-surface atoms are represented by spheres of white (H), red (O), black (C) and blue (N). Molecules are similarly represented as collections of such atoms in appropriate structures. H$_2$ molecules are colored yellow for ease of identification.}
    \label{figice}
\end{figure}

\section{Results}\label{results}

Fig. \ref{figice} shows ray-traced images of the cross-sections through the large porous and non-porous grains, at selected points in the model runs. From left to right, the images depict the growth of the ice mantles at comparable moments (based on the degree of gas-phase depletion of heavy species). For the non-porous grain, the material from the gas phase builds up relatively uniformly over the entire surface. There is also no indication of empty pockets. Based on the results of the hydrogen/oxygen system modeled by \citet{garrod13a}, the ice produced on dust grains under dark-cloud physical conditions should be non-porous, as heavy atoms accreted from the gas phase would have ample time to diffuse on the surface, finding and filling any nascent surface inhomogeneities, which tend to manifest initially as stronger binding sites due to their local curvature.

For the porous dust grains, after a sufficient amount of ice mantle has been built up (interval 100), the surface appears to be relatively uniform as well. However, looking below the surface, there are some differences. Firstly, rather than filling up the pores in the grain material, the ice fills them only partially, with some porosity remaining as the ice builds up over the top, leaving empty pockets that are not present in the non-porous grains.  Secondly, there appears to be a collection of molecular hydrogen gathered in those remaining porous structures.

\begin{figure}[ht]
    \centering
    \includegraphics[width=0.99\textwidth]{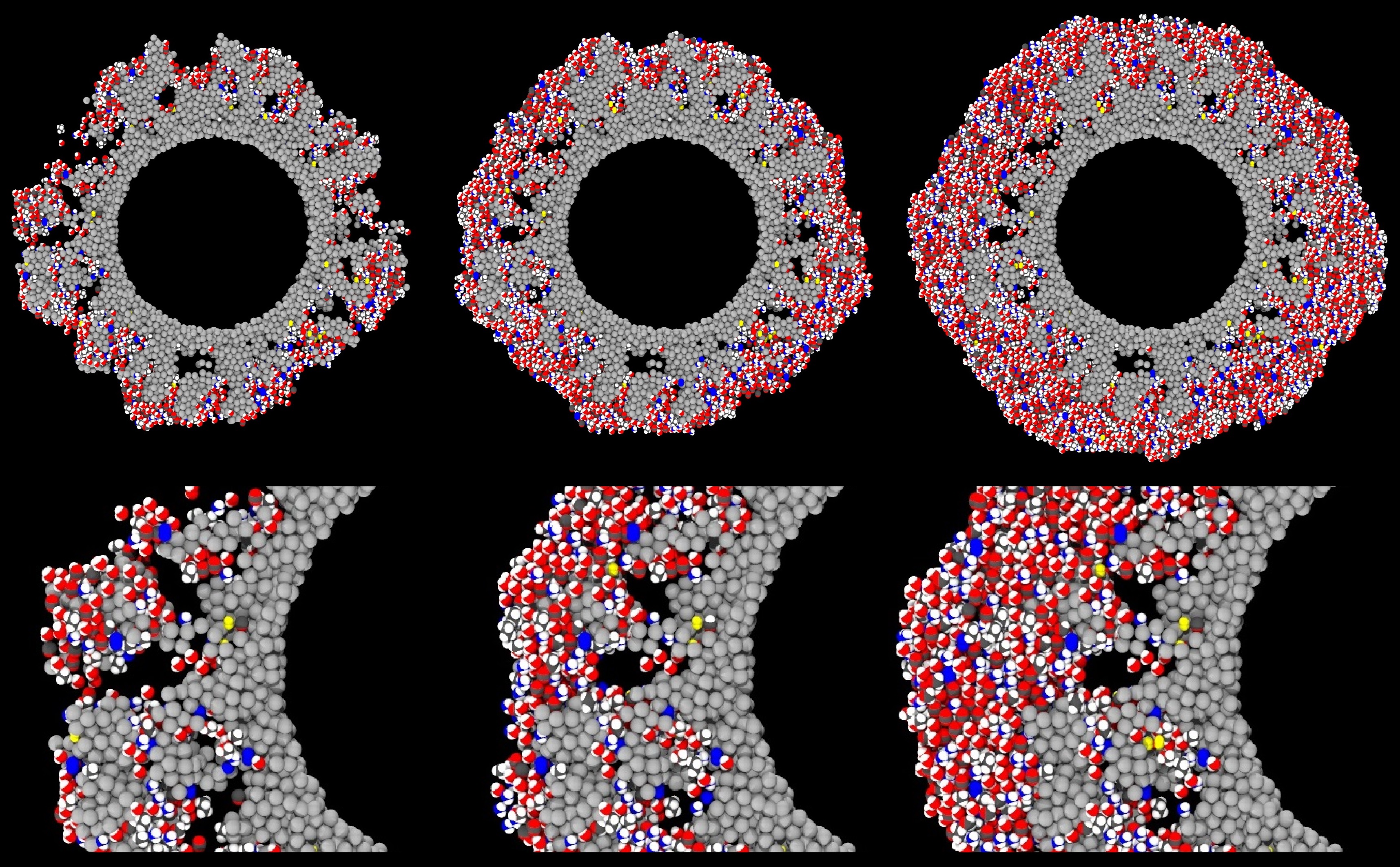}
    \caption{
    Ray-traced, cross-sectional images of the large (100 {\AA} radius) porous grain type at the early stages of ice build-up, corresponding to the grain shown in the bottom row of Fig \ref{figice}; here, the cross-section at a different angle is shown. Left to right: interval 25, interval 50, interval 100. The three upper panels show the full cross-section, while the lower panels show close-up images of the same small region of the grain at each of the three intervals. Gray spheres represent the carbon atoms composing the dust grain. Grain-surface atoms are represented by spheres of white (H), red (O), black (C) and blue (N). Molecules are similarly represented as collections of such atoms in appropriate structures. H$_2$ molecules are colored yellow for ease of identification.
    }
    \label{figzoom}
\end{figure}

Fig. \ref{figzoom} shows cross-sectional close-up images of the large porous grain at early times in the ice-mantle evolution. Porous structures are gradually filled, sometimes leaving voids in the ice. While much of the volume of the pore voids present in the original grain is filled, some of the bare grain surface remains and there are internal ice surfaces formed where the pores are not completely filled. H$_2$ molecules, marked in yellow, can also be seen more clearly in Fig. \ref{figzoom}; some are apparently bound directly to the grain material itself, while others are embedded in the ice. Notably, the H$_2$ molecules present at interval 25 (left-most images) seem mostly to retain their positions at later intervals, indicating that they are well fixed into those structures. It may be noted that the ices that form once the pores in the grain material are covered over are themselves non-porous; as found by \citet{garrod13a}, under dark-cloud conditions, oxygen and other atoms have long enough pre-reaction lifetimes before they react with the more mobile H atoms (producing e.g. water), that they may diffuse enough to find strong surface binding sites, leading to a smoothing of the ice surface. The porosity that is present in the ice by the end of the model runs is due to the initial presence of porosity in the grain material. The final grain/ice does not retain that same degree of porosity, as the ice partially fills the original voids, and then forms non-porous ice once those are fully enclosed.

\begin{table}[]
    \centering
    \caption{The total abundance of some notable ice species averaged over three runs. The final time of each model is also indicated, which corresponds to interval 1000 in each case.}
    \begin{tabular}{ l || r | r | r || r | r }
        & \multicolumn{3}{c||}{Non-Porous} & \multicolumn{2}{c}{Porous}\\
\hline
        Species & \multicolumn{1}{c|}{Small} & \multicolumn{1}{c|}{Medium} & \multicolumn{1}{c||}{Large} & \multicolumn{1}{c|}{Medium} & \multicolumn{1}{c}{Large} \\
                &  (34,885 yr) &  (56,417 yr) &  (71,851 yr) &  (66,636 yr) &  (83,484 yr)  \\
\hline
        H        &   7 &  32 &  62 &  43 &  92  \\
        H$_2$    &   31 &  214 &  463 &  363 &  895  \\
        C        &   47 &  69 &  193 &  98 &  200  \\
        O        &   42 &  75 &  160 &  100 &  177  \\
        N        &   84 &  170 &  507 &  250 &  522  \\
        OH       &   116 &  109 &  338 &  205 &  341  \\
        CO       &   3,002 &  10,699 &  23,180 &  10,739 &  22,943  \\
        N$_2$    &   10,420 &  36,876 &  82,028 &  38,565 &  82,163  \\
        O$_2$    &   182 &  467 &  1,319 &  609 &  1,339  \\
        CO$_2$   &   33,240 &  121,364 &  263,505 &  122,636 &  262,636  \\
        CH$_4$   &   906 &  4,208 &  8,466 &  3,879 &  8,907  \\
        NH$_3$   &   16,560 &  63,815 &  134,897 &  62,443 &  134,043  \\
        CH$_3$OH &   502 &  2,549 &  4,958 &  2,297 &  5,353  \\
        H$_2$O   &   67,187 &  247,960 &  533,161 &  249,904 &  534,938
    \end{tabular}
    \label{tab:abuns}
\end{table}

Table \ref{tab:abuns} shows the absolute number of atoms and molecules resident on the grain surfaces at the end of each model. Values shown are the mean final populations of the three random number-seed runs for each setup. The final time in each model is also indicated. Table \ref{tab:relativeabuns} shows the same population data given as a fraction of the total water on the grains, which allows an easier comparison between models. The porous grains take a little longer to reach their final state (i.e. the same amount of ice) as, compared with their non-porous equivalents, the porous grains present slightly smaller accretion cross sections.

Comparing between models, firstly we may consider the effect of grain size, by contrasting the relative abundances achieved in each of the three non-porous grain models. The abundance data suggest that variation in size has little effect on the relative abundances of the main ice constituents, such as CO, CO$_2$, and other stable molecules. CO reaches a population a little over 4\% of total water, while CO$_2$ is at around 49\%. In comparison to observational values toward background stars, the latter is a little high, while the former is a little low; observational lower and upper values are 9--67\% and 14--43\%, respectively \citep{boogert}. This may be explained by our use of static physical conditions throughout the model run, combined with the lack of gas-phase chemistry. The initial chemical abundance of CO is representative of a somewhat evolved gas-phase chemistry, while that of atomic oxygen is more appropriate to early times; this combination therefore leads to a somewhat higher degree of conversion of CO to CO$_2$ than observations would suggest.

The fractional abundances of volatile species on the grains do not scale so well with grain size; C, O, N and OH are seen to fall with increasing grain size. The absolute abundances of these species increase with greater grain radii, but not at the $r^3$ ratio seen for stable species. However, these values are likely reflective of the lifetime of such species against reaction with a newly-accreted H atom on the grain surface, which is more dependent on grain surface area than on total volume. The differences between grain-size models do not otherwise seem to affect the resultant chemical compositions of stable molecules.

The two porous grains may also be compared. Fractional abundances between the medium- and large-sized porous grains are again similar for stable species, while the atoms and radicals show somewhat less variation; indeed the fractional abundance of atomic H is identical between the two. We note also that the quantities of atomic H shown in the Table correspond largely to H trapped within the bulk ice; the total, instantaneous population of free (surface) H atoms on the grains in any of the models hovers around one.

Comparison of the abundances of each chemical species between the porous and non-porous models of the same grain size demonstrates an interesting outcome. Beginning with the medium grains, the final fractional abundances are fairly consistent with one another for most species, especially stable molecules. The one molecular abundance that stands out in comparison between porous and non-porous grains is that of molecular hydrogen, which is about two times greater in the porous cases than it is in the non-porous cases, by the end of the model runs. The absolute numbers of H$_2$ on the grains are on the order of hundreds, so the differences cannot be due to random small-number fluctuations. Consideration of the large porous and non-porous grains shows similar behavior for all species, while the discrepancy in the absolute abundance of H$_2$ is further exacerbated. Atomic H also shows a much stronger variation than the other atoms when comparing between porous and non-porous grains.

\begin{figure}[ht]
    \centering
    \includegraphics[width=0.6\textwidth]{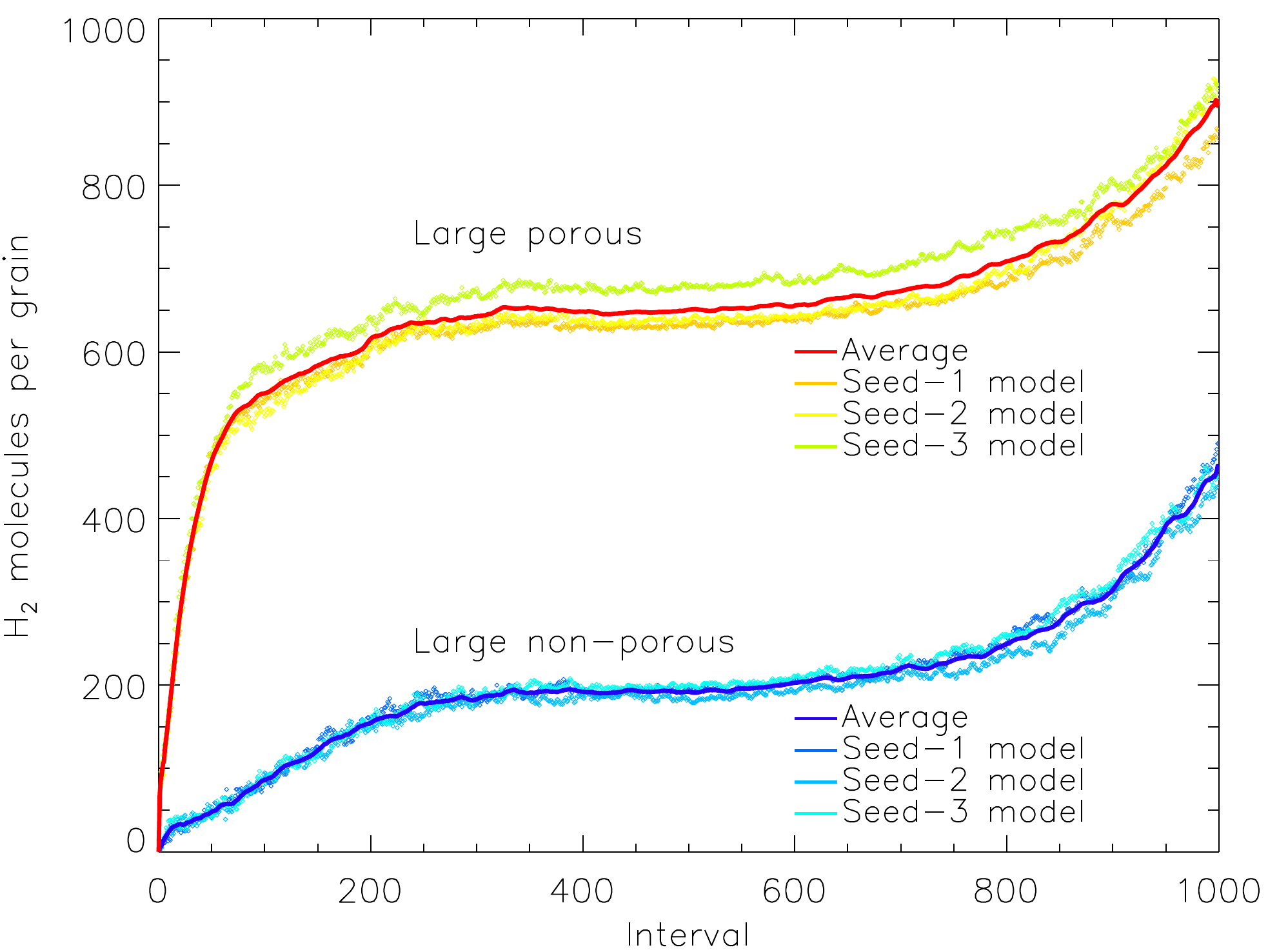}
    \caption{The amount of molecular hydrogen present on the grain versus relative time intervals, for the large porous and non-porous grain models. The solid lines show a moving average (of period 10) over the three runs for either grain type, with the data from the individual runs plotted as individual points. Interval 1000 corresponds to the end time of the models shown, which are shown individually for each model in Table \ref{tab:abuns}.}
    \label{fig2}
\end{figure}

\begin{figure}[ht]
    \centering
    \includegraphics[width=0.6\textwidth]{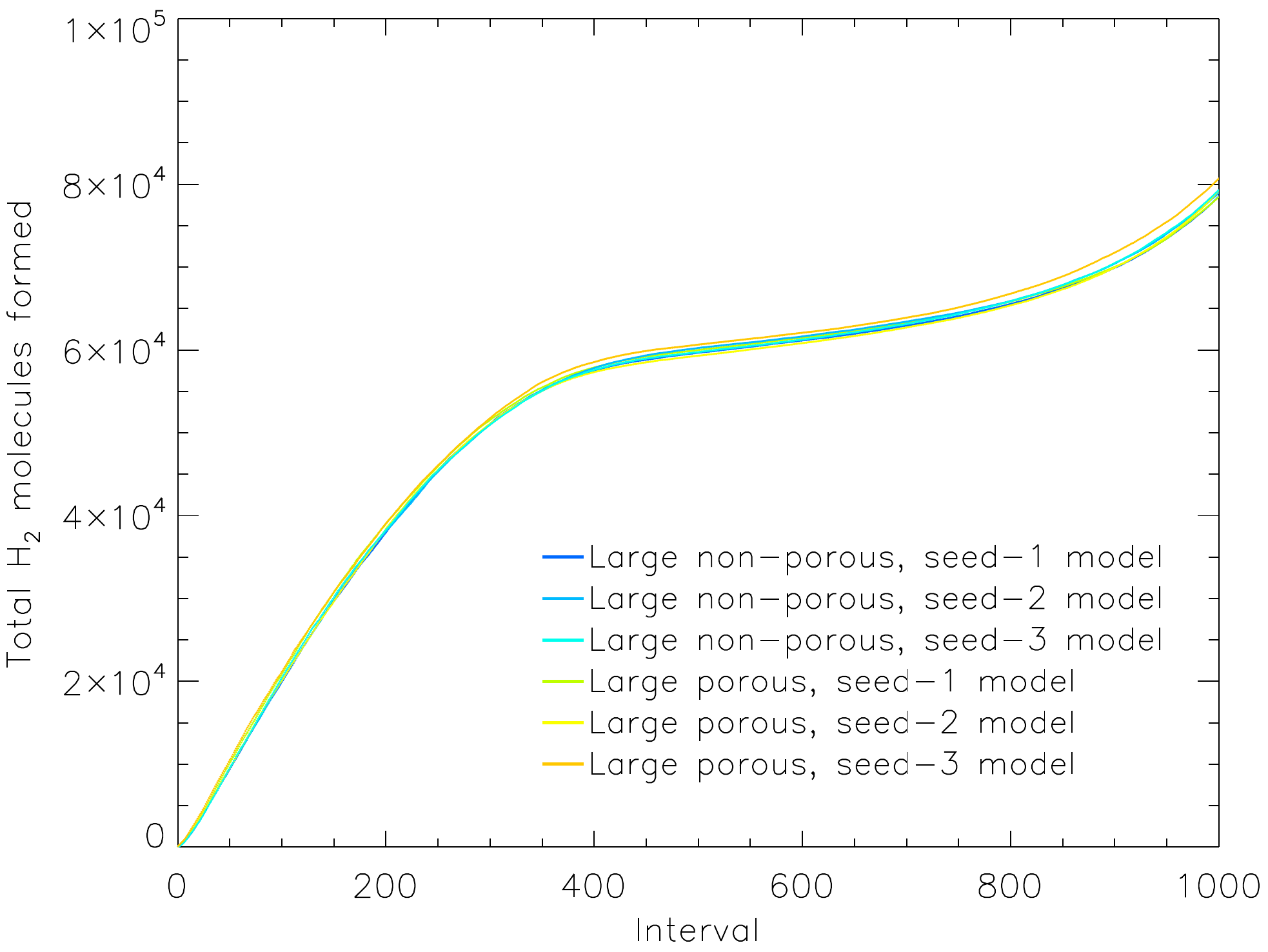}
    \caption{The total count of hydrogen molecules formed in the model up to a given moment, versus relative time intervals. All H$_2$ formed is shown, regardless of whether it is trapped in the ice, desorbed into the gas phase, or destroyed by chemical reactions. Data from the three runs for the large porous and non-porous grain models are shown. Interval 1000 corresponds to the end time of the models shown, which are shown individually for each model in Table \ref{tab:abuns}.}
    \label{fig3}
\end{figure}

In order to pinpoint the cause of this, the time-dependence of the H$_2$ population may be examined. Fig. \ref{fig2} shows the absolute populations of H$_2$ for the large porous and non-porous grains, averaged over the three random number-seed runs for each case; each of the individual runs is close to its mean curve. For most intervals, the porous and non-porous grains show almost identical trends, except that the porous-grain model has a very rapid up-tick at early times -- times at which the porous grain-surface is still exposed. Ignoring this offset, there is otherwise convergence in behavior after that initial jump. It is evident that the enhancement in H$_2$ population at those early stages becomes a permanent feature of the ice, as the offset is retained right until the end of the models. The growth in the H$_2$ populations toward the end of both sets of models is caused by the falling rates of accretion of heavy species from the gas phase, as their limited budgets run out, while H atoms remain plentiful throughout, increasing the probability of H$_2$ production (the same effect is seen in Fig. \ref{fig3}).

\begin{table}[]
    \centering
    \caption{The fractional abundance of some notable ice species relative to the total water ice, averaged over three runs.}
    \begin{tabular}{ l || r | r | r || r | r }
        & \multicolumn{3}{c||}{Non-Porous} & \multicolumn{2}{c}{Porous}\\
\hline
        Species & \multicolumn{1}{c|}{Small} & \multicolumn{1}{c|}{Medium} & \multicolumn{1}{c||}{Large} & \multicolumn{1}{c|}{Medium} & \multicolumn{1}{c}{Large} \\
\hline
        H        &   1.09e-4 &  1.24e-4 &  1.16e-4 &  1.72e-4 &  1.72e-4  \\
        H$_2$    &   4.57e-4 &  7.09e-4 &  8.69e-4 &  1.45e-3 &  1.67e-3  \\
        C        &   6.99e-4 &  4.32e-4 &  3.63e-4 &  3.92e-4 &  3.74e-4  \\
        O        &   6.29e-4 &  4.71e-4 &  3.00e-4 &  4.00e-4 &  3.30e-4  \\
        N        &   1.25e-3 &  1.02e-3 &  9.52e-4 &  9.99e-4 &  9.76e-4  \\
        OH       &   1.72e-3 &  8.16e-4 &  6.33e-4 &  8.19e-4 &  6.38e-4  \\
        CO       &   4.47e-2 &  4.36e-2 &  4.35e-2 &  4.30e-2 &  4.29e-2  \\
        N$_2$    &   1.55e-1 &  1.54e-1 &  1.54e-1 &  1.54e-1 &  1.54e-1  \\
        O$_2$    &   2.71e-3 &  2.34e-3 &  2.47e-3 &  2.44e-3 &  2.50e-3  \\
        CO$_2$   &   4.95e-1 &  4.91e-1 &  4.94e-1 &  4.91e-1 &  4.91e-1  \\
        CH$_4$   &   1.35e-2 &  1.46e-2 &  1.59e-2 &  1.55e-2 &  1.67e-2  \\
        NH$_3$   &   2.46e-1 &  2.50e-1 &  2.53e-1 &  2.50e-1 &  2.51e-1  \\
        CH$_3$OH &   7.47e-3 &  8.52e-3 &  9.30e-3 &  9.19e-3 &  1.00e-2
    \end{tabular}
    \label{tab:relativeabuns}
\end{table}

There are several possible explanations for the larger population of H$_2$ on the porous grains. We propose the following as the most likely: (i) the porous surface enhances the overall conversion of H into H$_2$, with mobile H$_2$ formed in the pores diffusing away to be incorporated into the general bulk ices; (ii) the porous surface enhances production rates locally within the pores, leading to the build-up of H$_2$ where it is formed; (iii) H$_2$ that is formed anywhere on the grain surface may diffuse into the pores, where it becomes trapped.

Fig. \ref{fig3} shows the total number of H$_2$ molecules formed throughout each model; the plot shows all of the three runs for both the large porous and large non-porous grains. The overall production of H$_2$ is extremely similar between models and grain sizes, including at the early times when the effects of the porosity manifest. Any difference may be accounted for by slight variations in the sizes of the porous and non-porous dust grains. We can therefore rule out a difference in the production efficiency of H$_2$ on the porous grain as the cause of its greater H$_2$ retention.

\begin{figure}[ht]
    \centering
    \includegraphics[width=0.6\textwidth]{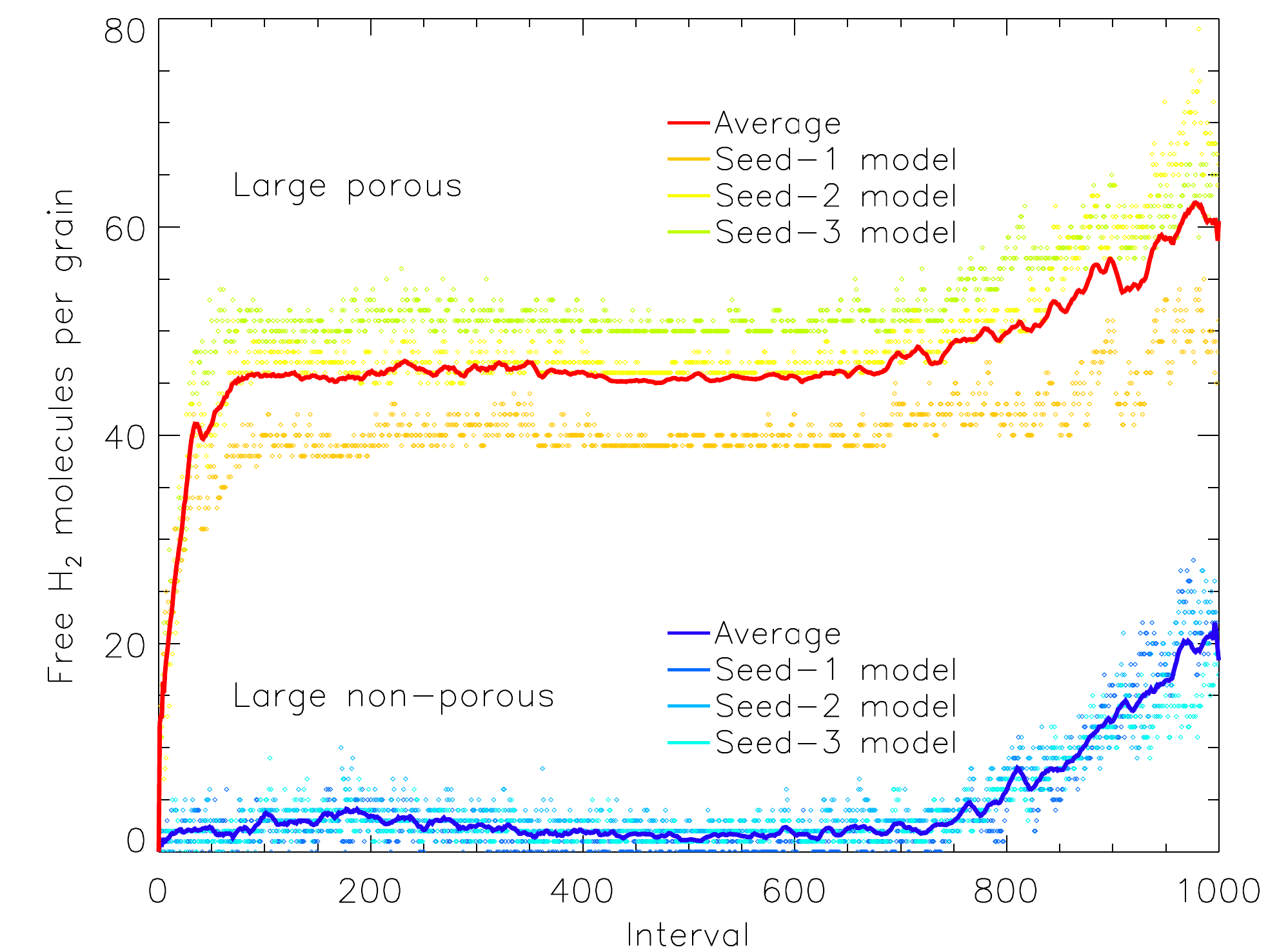}
    \caption{The amount of free molecular hydrogen present on the grain plotted versus relative time interval, for the large porous and non-porous grain models. Free species are able to undergo surface diffusion and desorption, but may reside on enclosed surfaces within pores. The solid lines show a moving average (of period 10) over the three runs for either grain type, with the data from the individual runs plotted as individual points. Interval 1000 corresponds to the end time of the models shown.}
    \label{fig4}
\end{figure}

It is instructive also to consider the amount of ``free'' H$_2$ throughout each model, shown in Fig. \ref{fig4} (large grains). This is a measure of the number of H$_2$ molecules that are capable of undergoing diffusion or desorption, at whatever rate, and which may therefore be classified specifically as surface species. Molecules that are not ``free'' are necessarily trapped under other species. The behavior of the free H$_2$ is qualitatively similar to that of the total H$_2$ population, and it also grows strongly at early times in the porous grain models. The absolute number of free H$_2$ molecules in the porous model reaches a peak of around 45 until late times, which is a factor $\sim$15 lower than the total H$_2$ on the grain at that time. Over the same period in the non-porous case, there is only a few free H$_2$ molecules at any one time, corresponding to roughly one in fifty of the total H$_2$ on that grain.

\begin{figure}[ht]
    \centering
    \includegraphics[width=0.4\textwidth]{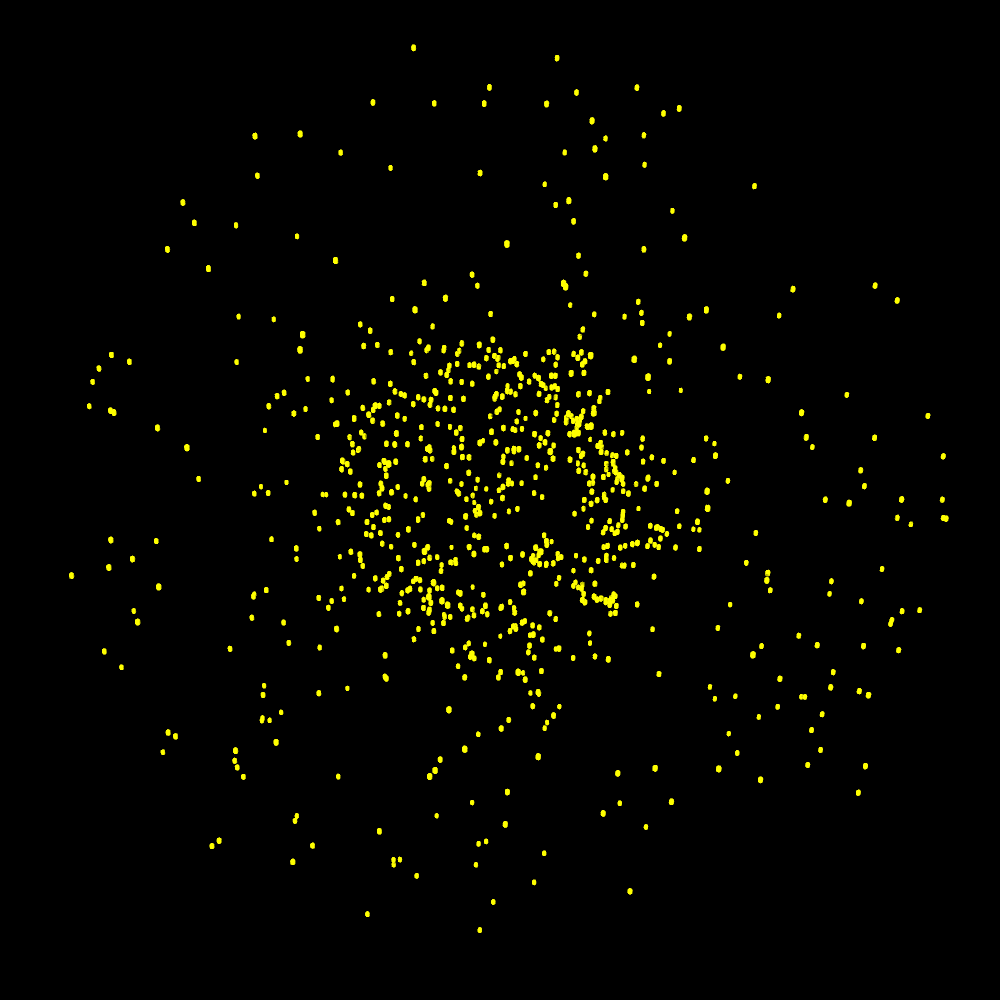}
    \caption{Ray-traced image of the large porous grain type at the final stage of ice build-up showing only the H$_2$ molecules and excluding the grain and all other ice species.}
    \label{figh2}
\end{figure}

Clearly, the much greater absolute numbers of surface H$_2$ on the porous grain indicates that those particular molecules at least are stored on the surfaces of the open porous structures that become closed off to outside chemical influence by around interval 100. With such a build-up in surface H$_2$ abundances in the pores, it is natural that some H$_2$ should become incorporated into the ice mantles that form specifically within the pores (i.e. no longer free). Fig. \ref{figh2} shows an image of the H$_2$ present on the large porous grain at the end of one of the runs, with no other atoms, molecules, or grain material depicted. The image includes all H$_2$ molecules present in three dimensions, not just a cross-section. The distribution of H$_2$ clearly favors the pore structures, although some H$_2$ is still present in the outer mantles.

We again consider the feature seen in Figs. \ref{fig2}--\ref{fig4} in which the amount of H$_2$ (whether the amount on the grain, the number of free molecules, or the total formed) increases over the last few hundred intervals. This effect is the result of the run-down in the gas-phase budget of species other than hydrogen, i.e. the depletion of those species. While, for example, atomic O becomes less and less abundant in the gas phase over time, due to its adsorption and incorporation into the ice mantle, the amount of gas-phase H stays the same; this means that for any one adsorbed H atom, its surface reaction partner is increasingly likely to be another H atom, which drives up the H$_2$ production rate. Meanwhile, the time required for a fixed number of gas-phase atoms (or CO) to be adsorbed from the gas phase also rises, due to their smaller remaining budgets. The amount of time associated with each interval therefore lengthens (see Sec. 2.2), making the increase in H$_2$ molecules appear more dramatic.

\begin{figure}[ht]
    \centering
    \includegraphics[width=0.6\textwidth]{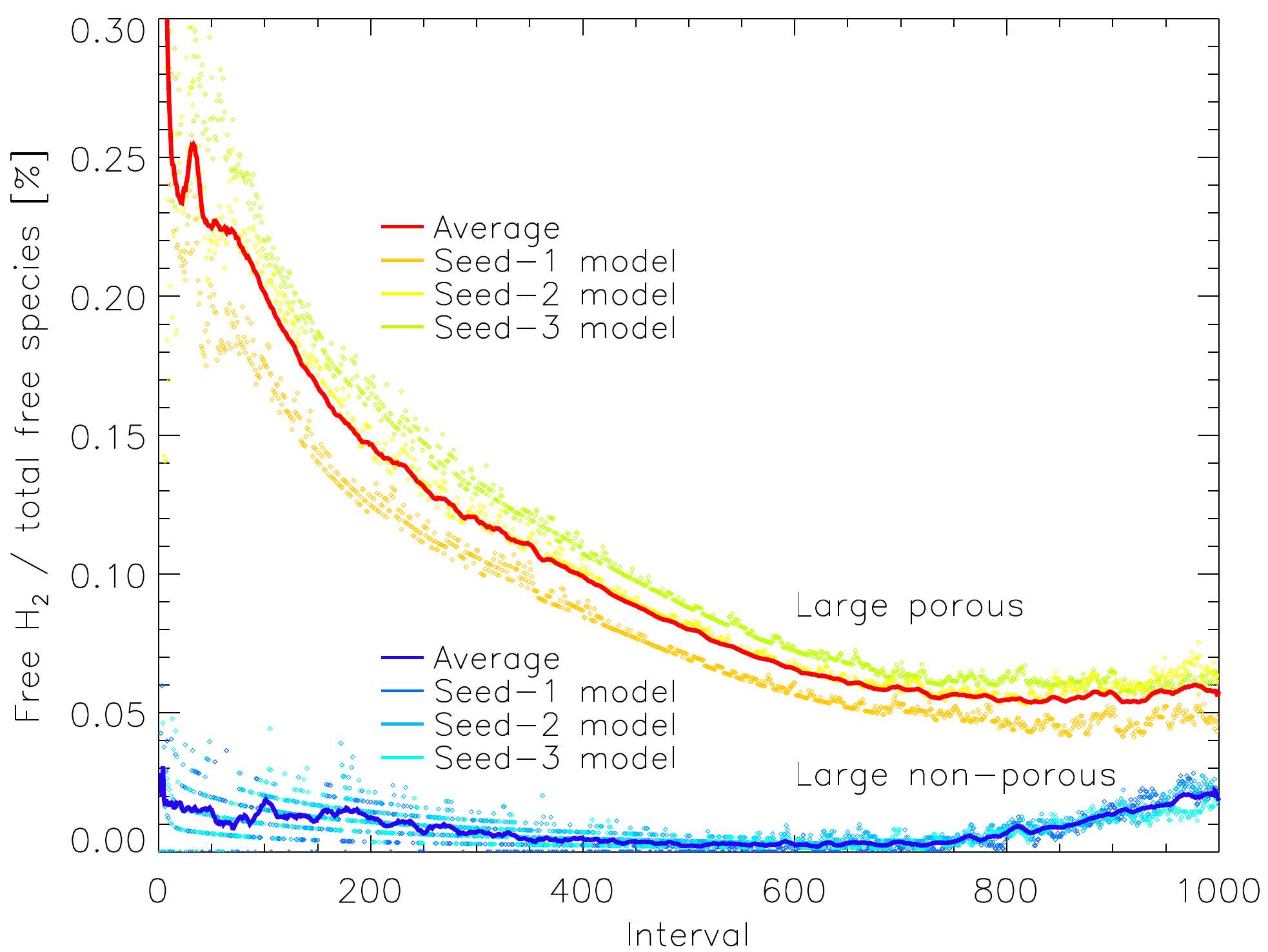}
    \caption{The percentage of free molecular hydrogen present on the grain as a function of all free species, plotted versus relative time interval for the large porous and non-porous grain models. Free species are able to undergo surface diffusion and desorption, but may reside on enclosed surfaces within pores. The solid lines show a moving average (of period 10) over the three runs for either grain type, with the data from the individual runs plotted as individual points. Interval 1000 corresponds to the end time of the models shown.}
    \label{percent_free_H2}
\end{figure}

Fig. \ref{percent_free_H2} shows the amount of free H$_2$ on the grain, just as in Fig. \ref{fig4}, but is shown here as a percentage of the total population of free species. Note that the number of free species in this model corresponds approximately to the number of species (atoms or molecules) present in the surface of the ice or on the surface of the grain (during the early stages before a mantle has built up). This count of surface species includes not only the outer surface but any enclosed surfaces that remain within the pores. As a fraction of the total, we see initially that the difference between the porous and non-porous grains is quite stark, as the pores in the grain material retain a substantial amount of H$_2$. The difference gently declines for later intervals for two reasons: (i) much of the porous structure within the grain material becomes filled up and smoothed out, so that some of the surface H$_2$ trapped there is locked into the bulk ice that resides in the grain pores, and (ii) the outer surface of the ice mantle grows, so that the enclosed surfaces that remain deep down in the grain/ice account for a smaller fraction of the total surface available. By the end of the two runs, the percentage of surface coverage by H$_2$ is similar (within a factor of 2); the difference that remains nevertheless corresponds to the H$_2$ that is trapped within porous structures, residing on their enclosed surfaces and unable to escape. Plotted as a percentage of total coverage, the apparently dramatic rise in free H$_2$ (seen in Fig. \ref{fig4}) is far more subdued in both models.

\section{Discussion}

Aside from some small variations in fractional abundances, neither the grain size nor porosity appear to have a major effect on the grain-surface ice composition, except in the case of H$_2$. Atomic H is also retained to a somewhat higher degree on the porous grains, but the absolute numbers are relatively small compared with H$_2$.

Porous grains retain molecular hydrogen uniquely, either in the ice mantles that form within the pores, or on the surfaces of the pore mantles, in cases where voids exist within the icy pores (having been partially filled and then closed over by ice). The porosity does not induce any greater production of H$_2$ (or other stable molecules) on the grains; rather it acts simply to trap those molecules. However, even if the total H$_2$ produced is essentially unchanged, it is a different question to ask whether the presence of pores affects where on the grain that fixed amount of H$_2$ is formed. Trapping of atomic hydrogen on pore surfaces could lead to local production of H$_2$ that may likewise become trapped; alternatively, H$_2$ produced elsewhere on the grain could also become trapped, following surface diffusion into the pores. Both are plausible explanations for the excess of trapped H$_2$ in the pores and pore ices.

As described by \citet{garrod13a} in the case of porous ices produced on smooth grains, the reason molecular hydrogen becomes trapped in pores at all has to do with the effect of surface structure on binding energies. The surface of a smooth grain is flatter; thus, fewer grain-surface binding partners maintain close proximity to a given hydrogen molecule at any time. However, the interior of the pores is more irregular and concave on very small scales, allowing more grain atoms or bulk-ice molecules to contribute to the binding energies and diffusion barriers experienced by a particle within that structure. The resulting stronger binding site allows for the molecular hydrogen to be held within the pore for long enough for other ice material to cover it up, or at least to seal over the pore before it is able to escape. The larger proportion of ``free'' H$_2$ molecules in the porous, versus non-porous, grain models indicates that the sealing of porous voids is a non-negligible effect, although the complete incorporation of excess H$_2$ into the pore ices is the primary cause of retention.

Atomic hydrogen is mobile on the grain surfaces, but upon entering a pore it may encounter a strong binding site and be retained there for a longer period than if it were in a weak binding site. The arrival of another H atom would convert it to H$_2$, allowing that molecule to be trapped in that site. It may be seen, though, that other molecules -- whose progenitor-atoms such as O, C and N are much less mobile on the grains -- can also fill up the pores; the pores are not filled solely or even predominantly with H$_2$. Also, the overall ice structure outside of the grain-pores does not become porous itself once the underlying pores are covered over. This is because the heavier atoms in the model are still sufficiently mobile to find strong binding sites for themselves, including those located within the pores, prior to being hydrogenated by atomic H (which leads to a stronger surface bond and much lower mobility, at which point that species becomes fixed). 

Why, then, should H$_2$ be uniquely affected by the presence of many strong binding sites within the pores? H$_2$ is unique also in being a mobile surface species that is largely unreactive. It may react indeed with OH, for example, but all of its few reactions have an activation energy barrier. On the other hand, atoms like H, O, C, and N, even though reasonably mobile, will react immediately upon meeting some other reactive species, ending their diffusion. Their diffusion path length on the surface is therefore limited by meetings with potential reaction partners, as well as by their somewhat slower diffusion rates (versus their main reaction partner, atomic H). An H$_2$ molecule, however, has the chance to sample a much larger fraction of the grain surface, seeking out strong binding sites, with its main loss route ultimately being desorption into the gas phase rather than reaction. As well as having a greater chance to find those strong sites, the lifetime against desorption of an H$_2$ molecule will be substantially increased if it does find one.

Given a large selection of strong binding sites on the grain, due to their high mobility H$_2$ molecules are most likely to fill them. The availability of a larger number of strong sites in the porous case therefore invites the retention of H$_2$. When strongly bound species are formed on top of the H$_2$, or in close quarters, that H$_2$ becomes trapped. The disappearance of strong binding sites as the ices build up and become more uniform in structure removes this effect.

While the numbers of H$_2$ being trapped in the present models are on the order of hundreds, larger grains would likely store proportionately more. Also, the model specifically leaves out the accretion of H$_2$ directly from the gas phase; all of the trapped H$_2$ in the models must first be formed on the grains through chemical reactions. In reality, grains would be under a constant flood of H$_2$ adsorption, even if the individual lifetimes of those molecules were short. We should expect that any strong binding sites in the grain pores would be rapidly and comprehensively filled by H$_2$ in that case. The potential for trapping could therefore be much higher in a more realistic simulation. Unfortunately, as the H$_2$ abundance in the gas phase is around 1000 times higher than the other species collectively, at current computational speeds those simulations would take years rather than weeks.

Similarly to molecular hydrogen, the trapping of other unreactive, volatile species such as helium could also be important. Interstellar He abundances are around 10\% with respect to total H (by number), and could therefore similarly flood the surface. Neon, as the next inert element, might also be sufficiently mobile on the grain surfaces to find strong binding sites and therefore experience trapping. If these or other inert species could be trapped in the deepest layers of the ices, close to the grain surfaces themselves, then they could be retained on the grains long into the star formation process, even under warm conditions where much of the ice material might have already desorbed. Such a mechanism may be contrasted with the ice annealing process explored by, e.g., \citet{bar-nun87}, whereby highly-porous water ice was used to trap Argon gas, with the pores being closed over through a diffusive creeping mechanism of the ice at the pore openings.

It should be noted that it is specifically the microstructure of the porous grains that is most important to the retention of H$_2$, or other volatiles similarly affected, in the models presented here. The coagulation of large, smooth grains would not likely produce such an effect. However, the presence of an icy coating on the grain pore surfaces may be just as effective as the bare grain itself in trapping H$_2$, so long as the build-up of the ice within those pores allows the underlying microporous and/or irregular structure to be retained.

The release of the chemical energy of surface reactions has been suggested as a mechanism for both the general heating of a dust grain and the non-thermal desorption of H$_2$ in particular \citep[see e.g. the review of][]{Wakelam17a}. Calculations of the heating effect of H$_2$ production on silicate grains through the addition of H atoms by \citet{NR14} suggested a strong influence on the temperatures of very small grains. Scaling their calculations to the size of our ``large'', 100~\AA~radius grains, the expected change in temperature would be less than 0.1~K per H$_2$ production event, with a time constant for the radiation of the excess energy of less than 1~$\mu$s, i.e much faster than the rate of H$_2$ production. The whole-grain effect of chemical heating (at least via H$_2$ formation) is therefore unlikely to have a significant influence on our models, although the somewhat smaller local structures produced in our porous grains could be more strongly affected. Chemical heating could affect the local structure of the ice also; however, \citet{Oba09} suggest that in the case of amorphous ice formation through surface chemical reactions, the structure is not influenced by this mechanism, in line with calculations by \citet{kouchi94}.

The chemical energy of reactions could allow H$_2$ formed on the grains to desorb through direct chemical desorption, although in our models the desorption process follows explicit trajectories, so that an H$_2$ molecule desorbed in this way could certainly be re-adsorbed. However, if the surface of the grain is indeed flooded with H$_2$ molecules (see above), the influence explicitly of the H$_2$ formation process should be far less important than the diffusion of H$_2$ into the pores. Indeed, the models presented here show that the H$_2$ does not need to be formed {\em in situ} within the pore structures to become trapped there.

The kinetic models used here sample only a small region of parameter space in terms of grain size, porosity and small-scale structure/roughness. The simulated dust-grains are constructed crudely to produce high degrees of porosity. A more detailed determination of the precise microstructures of interstellar dust grains is therefore necessary to constrain the proposed effect using chemical kinetic models. It is also the case that the carbon atoms that constitute the grain material in our treatment are not allowed to participate in any way in the surface reactions; we do not, for example, consider chemisorption in the production of H$_2$ or other species. As discussed by \citet{Wakelam17a}, reactions between physisorbed H atoms are likely to be the dominant H$_2$ formation mechanism under translucent and dark cloud conditions, while substantial water or other ice mantles are not expected to build up before visual extinctions greater than around 3 magnitudes (i.e. translucent) are achieved \citep{whittet}.

Our model also uses only a simple, static treatment of the physical conditions of dark clouds, ignoring the lower density and visual extinction that would pertain at early times, and which could lead to a somewhat higher dust temperature. The initial chemical conditions are also quite simple, with CO and H$_2$ assumed already to have formed in substantial quantities at the time the models begin. The lack of a gas-phase chemistry could also affect the results; it is possible, for example, that the gas-phase production of water would lead to some component of the surface water ice originating through direct deposition from the gas phase; such conditions would be a little closer to those frequently employed in the laboratory for the study of porous amorphous water ices, and might therefore lead to a higher degree of porosity in the ices themselves that would not be associated directly with the grain material's porosity. However, past dark-cloud gas-grain chemical models by our group \citep[e.g.][]{gp11} have not found any substantial contribution from gas-phase water in the build-up of the ice mantles, even under conditions of low extinction and elevated dust-grain temperatures.

The possibility of thermal desorption and re-deposition of water ice onto the dust grains in protoplanetary disks, however, might provide a regime in which ices of substantial porosity could be formed, through direct deposition rather than chemical formation on the grain surfaces. H$_2$ or other volatiles might be expected to be trapped in such structures also.

\section{Conclusion}

Our chemical kinetic models of interstellar grain-surface chemistry indicate that, given a highly porous dust grain, molecular hydrogen may be trapped in strong binding sites caused by the irregular surface structure of the pore surfaces. The coating of those pore surfaces with water and other common dust-grain ice mantle constituents locks the H$_2$ into the ice, although the irregular surface structures of the ice itself continue to allow trapping until the pores are completely filled and the grain/ice takes on a more regular, spherical morphology. The closing over of the pores, creating voids, is only a small part of the trapping effect, but a fraction of the trapped H$_2$ is found to be free on the surfaces of those enclosed voids. A full-scale treatment of H$_2$ adsorption onto the grains using this microscopic treatment might be expected to exhibit even greater degrees of trapping. The trapping effect could provide a means for other volatile and relatively inert species to be retained on the grains, enclosed in the deepest ice layers. Such an effect would allow chemical species trapped under interstellar cloud conditions to be preserved in the ices into much later stages in the star- and planet-formation process.

\section*{Author Contributions}

DAC ran computational models, analyzed the data, and co-wrote the manuscript. RTG wrote the {\em MIMICK} computational code, helped analyze the data and co-wrote the manuscript.

\section*{Acknowledgements}

We thank the anonymous referees for comments and suggestions that improved this paper.

\section*{Funding}
This work was funded by the NASA Astrophysics Research and Analysis program, through grant number NNX15AG07G.

\bibliographystyle{frontiersinSCNS_ENG_HUMS} 
\bibliography{references}

\end{document}